\documentclass[sigconf]{acmart}
\AtBeginDocument{%
  }

\copyrightyear{2025}
\acmYear{2025}
\setcopyright{cc}
\setcctype{by}
\acmConference[HILDA' 25]{Workshop on Human-In-the-Loop Data Analytics}{June 22--27, 2025}{Berlin, Germany}
\acmBooktitle{Workshop on Human-In-the-Loop Data Analytics (HILDA' 25), June 22--27, 2025, Berlin, Germany}\acmDOI{10.1145/3736733.3736736}
\acmISBN{979-8-4007-1959-2/2025/06}

\usepackage{array}
\usepackage{xspace}

\newcommand*{\ONION}{\texttt{ONION}\xspace}

\usepackage{enumitem}
\usepackage{colortbl}

\begin{document}

\title{ONION: A Multi-Layered Framework for Participatory ER Design}

 \author{Viktoriia Makovska}
 \affiliation{%
   \institution{Ukrainian Catholic University}
   \city{Lviv}
   \country{Ukraine}}
 \email{makovska.pn@ucu.edu.ua}

\author{George Fletcher}
\affiliation{%
 \institution{Eindhoven University of Technology}
 \city{Eindhoven}
 \country{Netherlands}}
 \email{g.h.l.fletcher@tue.nl}

\author{Julia Stoyanovich}
\affiliation{%
  \institution{New York University}
  \city{New York}
  \country{USA}}
  \email{stoyanovich@nyu.edu}

\begin{abstract}
   We present \ONION, a multi-layered framework for participatory Entity-Relationship (ER) modeling that integrates insights from design justice, participatory AI, and conceptual modeling. \ONION introduces a five-stage methodology: Observe, Nurture, Integrate, Optimize, Normalize. It supports progressive abstraction from unstructured stakeholder input to structured ER diagrams. 
  
  Our approach aims to reduce designer bias, promote inclusive participation, and increase transparency through the modeling process. We evaluate \ONION through real-world workshops focused on sociotechnical systems in Ukraine, highlighting how diverse stakeholder engagement leads to richer data models and deeper mutual understanding. Early results demonstrate \ONION’s potential to host diversity in early-stage data modeling. We conclude with lessons learned, limitations and challenges involved in scaling and refining the framework for broader adoption.
\end{abstract}

\maketitle

\section{Introduction}

\paragraph*{Motivation and Goals.}  Drawing closely on design and HCI disciplines \cite{costanza2020design,Feinberg17}, in this work we investigate the possibilities for collaborative participatory design of data models.  Our work springs from two motivations.  The first is the critical importance of data modeling in the study of data management systems.  Indeed,  data-driven systems, including databases, statistical models, machine learning (ML) models, and artificial intelligence (AI) systems, are fundamentally dependent on data models. For example, consider a simple spreadsheet that contains payroll data for a small start-up. The columns headers might include "Employee name", "Employee address", "Employee gender", "Employee department," and "Department location." This data reflects an implicit data model composed of entities (Employee and Department), their attributes (e.g., "name" and "street address"), and relationships (e.g., the link between Employee and Department). 
The design of such data models (e.g., manifested as entity-relationship (ER) models, knowledge graphs, taxonomies, or ontologies) has long been an important aspect of data management. Indeed, data modeling is a critical first step before designing and implementing information systems.

Our second motivation is the urgent need to identify, reflect on, and mitigate the broader impacts of data systems. To create useful and inclusive models, we turn to \emph{participatory design} (PD)---an established approach that involves those most affected by technology in co-creating solutions \cite{schuler1993participatory}. Given the central role of conceptual modeling in formalizing and visualizing complex data structures, it is essential to develop participatory design approaches for data models so they reflect the needs, values, and lived experiences of both users and impacted communities.  PD approaches have been applied in information systems design \cite{kyng1991designing}, user-centered conceptual modeling \cite{vasilakis2019involving},  design of domain-specific ontologies~\cite{ongenae2011participatory, ongenae2014ontology, kingsun2018cdom, hocker2020participatory}, and critical data design practices such as data feminism \cite{dignazio2020data}. However, to the best of our knowledge, PD has not been applied to the construction of ER models.  

\smallskip

{\em Contributions.} 
We present \ONION, our work in progress toward a robust methodology for implementing PD in the classical ER modeling language. We focus on ER because it is widely taught and captures core features common to many modeling languages, making \ONION easily adaptable to other contexts. Drawing on our practical experiences, we analyze the challenges and opportunities of PD, including different modes of participation, strategies for involving stakeholders, and ways to make the process more effective, inclusive, and beneficial to those most impacted.

\section{Background}
We next review related work on participatory design, including design justice, participatory AI, and participatory modeling. We define a \emph{stakeholder} as anyone affected by a system’s use or its side effects, including indirect impacts. Our goal is to understand how participatory approaches can support the design of useful, inclusive, and fair systems---an essential step toward participatory frameworks for ER modeling.
    
\subsection{Design Justice and Inclusive Systems}

User-centered design (UCD) is a framework that emphasizes designing systems around the needs, preferences, and behaviors of end users, involving them throughout the development process \citep{norman2013design,garrett2010elements}. UCD employs methods such as usability testing, user personas, and iterative prototyping to align design decisions with user feedback. UCD is often positioned as a best practice for ensuring usability and relevance across application domains.

However, UCD has also been critiqued for centering dominant user groups and overlooking broader questions of equity and justice. \citet{costanza2020design} argues that, despite its inclusive intent, UCD can reinforce existing power imbalances and exclude marginalized communities. UCD methods such as \emph{lead user innovation}, which draws on insights from highly skilled or expert users, and \emph{user personas}, which create fictional characters to represent typical users, are intended to model user needs. Yet these techniques often reflect dominant perspectives, risk perpetuating stereotypes, and fail to represent those most affected by inequitable system design.

Consequently, \citet{costanza2020design} advocates for \emph{design justice}, which emphasizes diverse design teams and the direct involvement of marginalized communities. The core principles of design justice are: \textbf{inclusion}—actively listening to and acting on community needs; \textbf{equity}—ensuring diverse representation among designers and users; \textbf{Accountability}—empowering those most affected to lead the design process; and \textbf{ownership}—granting communities material control over both the design process and its outcomes.

This work warns that participatory design can unintentionally reinforce existing power structures if not carefully implemented. For instance, designers might extract knowledge from communities without providing benefits in return. \citet{costanza2020design} recommends avoiding tool-centric approaches, being cautious of ``tech parachuters'' (outsiders who suggest solutions without understanding local contexts), balancing innovation with necessary maintenance, and learning from existing participatory design models.

\subsection{Participatory Methods in AI}

Recently several teams have greatly advanced our understanding of  participatory methods in AI.
Collectively these authors discuss the importance of involving stakeholders in AI development and highlight the challenges and complexities of doing so effectively.

\paragraph{Types of Participation and Their Challenges}
\citet{birhane2022power} highlight the value of participatory methods in making AI systems more transparent, accountable, and responsive to those directly affected. They identify three forms of participation: \emph{active}, where stakeholders are directly and essentially engaged; \emph{incidental}, where stakeholders are involved but not actively engaged; and \emph{mediated}, where participation occurs through technological means.

The authors raise several concerns about participatory methods. These include: \textbf{the lack of clear standards for evaluation}, with potential criteria such as reciprocity, reflexivity, empowerment, and duration; \textbf{uncertainty about the goals of participatory tools}; and \textbf{the limitations of participatory methods}, including the risk of replacing democratic governance, corporate co-optation, conflating participation with inclusion, and challenges related to participant expertise and incentives.

Similarly, \citet{sloane2022participation} identify three forms of participation: \textbf{participation as work}, where stakeholders contribute labor to the design process; \textbf{participation as consultation}, where stakeholders provide input without decision-making power; and \textbf{participation as justice}, where participation is used to promote fairness and address power imbalances.

\citet{sloane2022participation} warn of ``participation washing,'' where participatory design is used superficially to satisfy legal or public relations requirements without meaningful stakeholder involvement. Hidden participation, where stakeholders are unaware of their involvement, can lead to misinterpretation of their influence and lack of compensation, reinforcing biases and power imbalances.

Both works highlight the potential pitfalls of participatory methods when not implemented genuinely. They emphasize the need for transparency, proper recognition of stakeholders' contributions, and aligning participation with clear, meaningful goals.

\paragraph{Addressing Power Imbalances}
\citet{suresh2024participation} examine the challenges of applying participatory methods to large-scale AI models. They introduce the concept of the ``participatory ceiling'' to describe structural limits on stakeholder involvement, driven by persistent power imbalances.

To address these limitations, \citet{suresh2024participation} propose a multilayered participation model: the \textbf{surface layer}, focused on task-specific input; the \textbf{subfloor layer}, addressing domain-specific needs and contexts; and the \textbf{foundation layer}, concerning general, context-independent aspects of AI models. Although their framework offers a structured approach to stakeholder involvement, the authors highlight persistent challenges with accountability, transparency, and fostering meaningful participation at each level—echoing concerns raised by \citet{birhane2022power} and \citet{sloane2022participation} about the practical limits of participatory methods.

\subsection{Participatory Modeling}
\citet{quimby2022participatory}
presents participatory modeling (PM) as a social science method for creating models of complex systems with direct stakeholder involvement. PM shares similarities with participatory design but focuses on modeling environmental or social systems.
The authors identify five key stages in PM.

    \smallskip
    \noindent {\em Selecting Roles and Sites.} 
In this step researchers answer essential questions: Who will participate? Who will facilitate? Where will the process take place? Sites should be accessible and inclusive, avoiding any barriers to participation. Two main roles are assigned: the facilitator, who guides discussions, and the modeler, who builds the model based on input.
    
    \smallskip
    \noindent {\em Identifying and Recruiting Participants.} 
Participants are selected based on their knowledge of or interest in the issue. Groups should be diverse, representing different perspectives within the community. Participants may come from various social or hierarchical backgrounds but must be connected to the community at hand.
    
    \smallskip
    \noindent {\em Formulating the Model.} 
Researchers and participants work together to set the model's goals and understand how different parts of the system interact. This can be done through interviews, surveys, or group discussions, allowing everyone to contribute.
    
    \smallskip
    \noindent {\em Conceptualizing and Building the Model.} 
During this stage, the actual model is created. This can involve using tools like maps, computer programs, or role-playing games. The participants continue to play a key role throughout this process, ensuring the model reflects their collective knowledge and priorities.

    \smallskip
    \noindent {\em Validating the Model.}
Once the model is built, it is tested for accuracy and relevance. Internal validity ensures that the voices of the community are represented accurately, while external validity checks how well the model fits within the broader context.
    
    \smallskip

\citet{quimby2022participatory} also highlight that PM is a political process. It has the potential to influence stakeholder perceptions and dynamics through communication. To avoid power imbalances, teams must be carefully formed. The earlier stakeholders are involved, the greater their influence on the overall outcome. 

\subsection{Conceptual modeling}
The process of conceptual data modeling is fundamental to any data modeling project, typically involving three main stages: identifying requirements, designing solutions, and evaluating those solutions  \cite{simsion2005data}. This is an iterative process where initial drafts lead to further questions, which in turn uncover additional requirements. The aim is to create a conceptual data model that fully addresses the needs of the application rather than searching for a single ``correct'' solution. The logical design phase in database development is typically portrayed as a designer-led activity focused on requirement gathering and conceptual modeling, with limited attention to the active participation of stakeholders \cite{silberschatz2020}.

Conceptual modeling is widely recognized as a design process \cite{Feinberg17,simsion}. Earlier works \cite{group_tooks_data_model_1997, group_tool_data_modeling_1997} focused on collaborative tools and meeting methods to support non-analyst participation in conceptual data modeling sessions, with an emphasis on business efficiency and task distribution. These efforts prioritized tool-centric approaches aimed at improving productivity, rather than participatory design as a justice-oriented practice.

Further, to the best of our knowledge, no explicit participatory design (PD) framework exists for entity-relationship (ER) models. While several papers explore participatory or collaborative approaches to \textit{conceptual schema} or \textit{data model} development, none focus specifically on ER as a formalism.
Indeed, \citet{ram1998collaborative} discusses collaborative conceptual schema tools that may support ER-like tasks but do not reference ER explicitly. Earlier group modeling tools \cite{group_tooks_data_model_1997, group_tool_data_modeling_1997, dean1994technological, lee2000cold} support participatory data modeling in business settings, but do not incorporate ER methods. Participatory ontology engineering frameworks \citep{ongenae2011participatory, ongenae2014ontology, kingsun2018cdom, hocker2020participatory} enable co-design of schema-like structures using formalisms such as OWL or RDF, but do not address ER-specific practices.
Finally, none of these earlier approaches 
focus on building safe collaboration spaces, building on recent progress in design justice and inclusive systems.

\subsection{Summary}

The reviewed literature underscores the importance of genuine stakeholder involvement in designing useful, just, and inclusive systems. Common themes include the need to avoid shallow participation, address power imbalances, and confront implementation challenges---without consensus on a single effective framework.

Most reviewed works define stakeholders as human actors directly affected by a system. We extend this view to include any entity—human or non-human—directly or indirectly impacted, including algorithmic agents. As \citet{choi2020situated} argue, recognizing algorithmic entities as stakeholders expands the scope of participation and reflects the complex interdependence within socio-technical systems. This perspective informs \ONION’s design, enabling it to account for diverse forms of influence and agency.

\ONION is grounded in the principle that authentic participation requires meaningful engagement at every stage of the modeling process. This includes addressing power dynamics, clarifying how input shapes outcomes, and developing clear evaluation criteria. Crucially, \ONION adapts participatory methods to the specific affordances and constraints of ER modeling.

The literature offers valuable critiques of current participatory design (PD) approaches. Building on these insights, our goal is to develop frameworks that retain the benefits of stakeholder participation while addressing its pitfalls—advancing effective, context-sensitive methods for the participatory design of ER models.

\section{\ONION: participatory approach for ER design}
\label{sec:framework}

We propose a participatory methodology to establish a foundation for a more inclusive data modeling environment from the earliest stages of system design, thus helping to mitigate biases within data models from the start. In this section we present derived methodology and its stages along with additional comments on roles and sites that shapes the form of the \ONION workshop.

\subsection{Derived methodology}
We name our approach \ONION as an abbreviation that describes all stages of our framework: Observe, Nurture, Integrate, Optimize, Normalize. The whole process is about going from a general and open system to a structured and closed system, which is necessary in order to translate the dynamic and fluent real world into a structured technological representation. This transformation mirrors the progressive abstraction process discussed in design theory, where a system evolves from high-level purpose and abstract function toward specific structures and physical forms \cite{gero1990design}.

The process must be fully participatory, involving stakeholders from the very beginning. When a relevant result is achieved, the process progresses to the next stage. However, if the result is deemed not relevant, the process goes back to the previous step, incorporating the necessary adjustments to refine the outcome. In some cases, this iterative approach may even require revisiting the initial step, ensuring that the design remains aligned with the evolving needs and inputs of all participants.

\begin{figure}
    \centering    \includegraphics[width=0.8\linewidth]{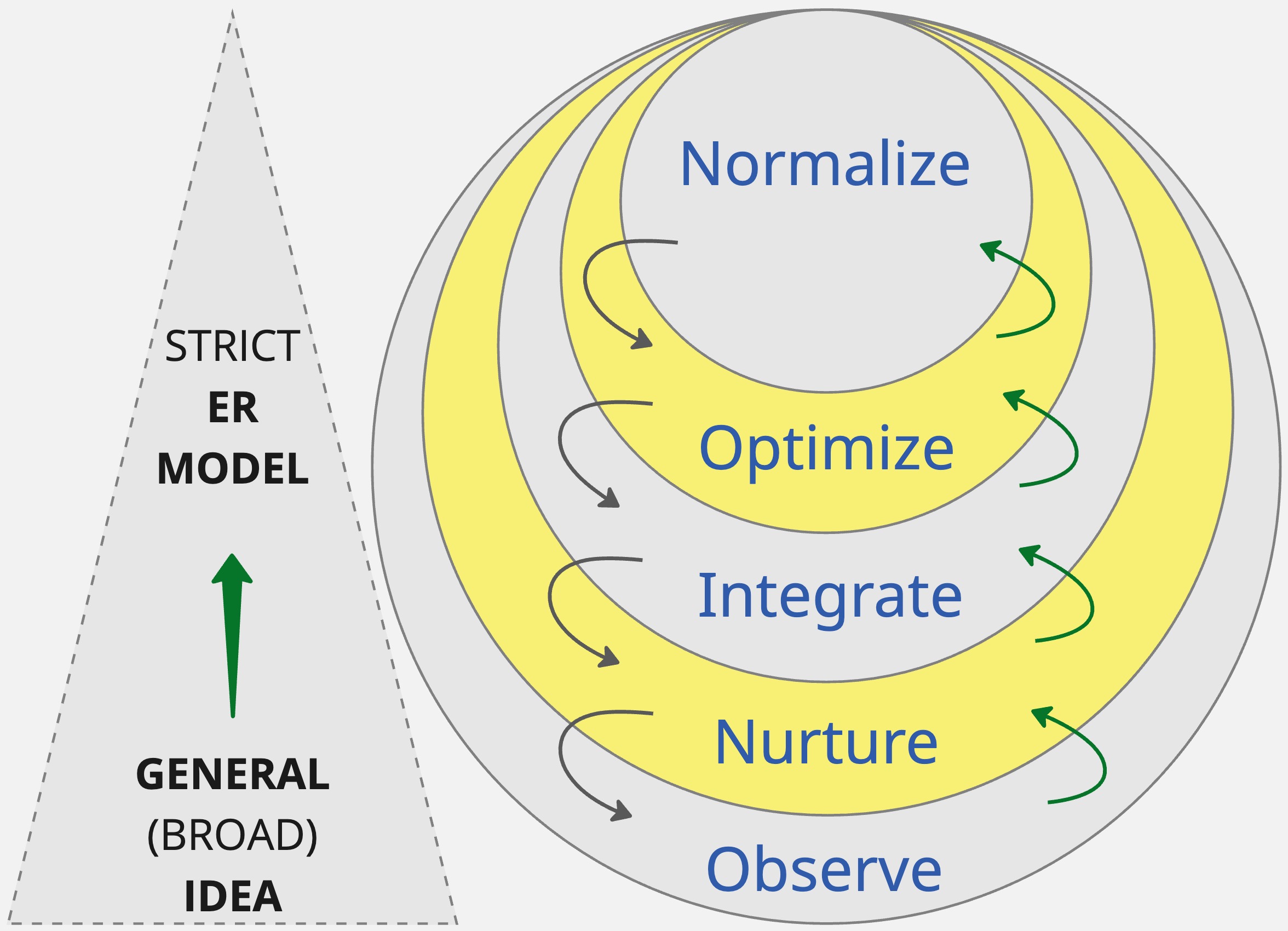}
    \caption{\ONION frame}
    \label{fig:refined-onion-frame}
\end{figure}

The concept of transitioning from an open-world approach to a closed system is highlighted on Fig. \ref{fig:refined-onion-frame} where green arrows indicate forward direction and gray arrows represent possible step back for stage refinement. Observe stage represents a broad and unrestricted world of ideas and that is why it is an outer layer of the frame. From this open space, we gradually shape the system with input from the participants, peeling of unrelated parts. In the early stages, participants are encouraged to share their thoughts freely, with no restrictions on form or text length. Each subsequent step narrows the focus, filtering out elements and highlighting the unique contributions and perspectives.

Ultimately, the final stage results in a strict and closed model, where every detail is precisely defined, marking a complete contrast to the open beginnings.

\subsection{Roles and Site}\hfill\
The \ONION framework is a participatory process where we need to define roles of participants to make it smooth and fluent. We suggest three key roles which are not limited and may be more granular in more complex cases: 
\begin{itemize} 
    \item \textbf{Participants}: Stakeholders engaged in questionnaires, collaborative activities, and internal validation. 
    \item \textbf{Experts}: Domain specialists responsible for synthesizing models and performing technical reviews. 
    \item \textbf{Facilitators}: Organizers who guide the process, recruit participants, and ensure contextual relevance. 
\end{itemize} 
The site---physical or virtual---is selected based on participant accessibility and project needs. Different settings may require additional participant training to ensure that everyone involved is familiar with the online platform being used or with the application of physical materials for visualization.

\subsection{Framework stages}
\paragraph{Observe.} This phase begins the abstraction process by gathering open-ended input from participants to identify initial needs and objectives. Stakeholders offer unstructured, free-form responses, helping surface real-world system understanding. This ``zero-assumption'' stage shapes the goals for subsequent design phases.

\paragraph{Nurture.} Here, collaborative modeling begins. All participants gather online or offline and are instructed about the workshop scenario and tools, and introduced to each other. In this stage, shared visualization options are used to bridge everyday language with conceptual modeling. Through guided activities, such as storytelling and schema sketching, they co-create narratives that reflect system needs. Participants are free to select the pace and form of visualization, while facilitators are in charge of conflict resolution and the overall flow of the discussion.

\paragraph{Integrate.} Experts synthesize participant input into a unified model. They identify entities, properties, and relationships, and construct an initial ER model. This stage represents a midpoint in the abstraction process, where informal insights are formalized.

\paragraph{Optimize.} Participants validate the draft model through feedback. This internal review ensures that technical interpretations still align with participants’ original intentions. Minor adjustments are made collaboratively to refine the structure.

\paragraph{Normalize.} In this final stage, the collaboratively constructed model is translated into a standard ER model. This represents the most structured layer, ready for technical implementation. An external technical expert review ensures the model is valid, complete, and suitable for development.\footnote{We use the term ``normalize'' differently than in conventional relational schema design.}
\section{Practical Investigation}
\label{sec:practical}

To evaluate \ONION in real-world conditions, we conducted three participatory modeling workshops aimed at capturing diverse stakeholder perspectives and transforming them into structured data representations. These case studies served as a testing ground for \ONION’s multi-layered framework, demonstrating how open dialogue and iterative refinement can lead to the co-creation of early-stage data models. Each workshop explored a different sociotechnical system and was adapted to suit the format, context, and participant needs. The following sections detail research questions, the workshop structure and outcomes from the Donor UA and Volunteers Lab case studies.

\paragraph{Research Questions}
Three main research questions shaped the workshop and the feedback process:
\begin{itemize}
    \item[RQ1:] Is \ONION usable and applicable? 
    \begin{itemize}
        \item RQ1.1: Is \ONION easy to understand and apply?
        \item RQ1.2: Does \ONION give participants freedom and possibility to express their ideas? 
        \item RQ1.3:Does \ONION reflect all stakeholder perspectives?
    \end{itemize}
    \item[RQ2:] Can \ONION help in designing ER models? 
        \begin{itemize}
        \item RQ2.1: Can \ONION produce a usable and technically correct ER model?
    \end{itemize}
    \item[RQ3:] Can \ONION support an inclusive data modeling environment?
    \begin{itemize}
        \item RQ3.1: Does the produced ER model capture vital information and contributions from participants?
        \item RQ3.2: Does the produced ER model mitigate any bias to any extent?
        \item RQ3.3: Do participants experience a change in their understanding of a given topic or a case study?
    \end{itemize}
\end{itemize}
We address these RQs in the Discussion section, where we give an overview of the early results and participant feedback.

\subsection{Structure and Process}
Before the workshop, we conducted an introductory session to communicate with participants and provide project information in textual form. Each workshop began with onboarding and clarification of objectives. Participants were then introduced to concepts, entities, characteristics, and relationships through facilitator guidance and open discussion. The facilitator maintained a low-bias environment by minimizing intervention, stepping in only for clarification or mediation.

Artifacts created by participants, such as story diagrams and conceptual sketches, were validated at the end of each session through structured reflection. One such approach was the ``information path'' or ``entity path,'' which ensured that tracing an entity through all stages of the model preserved system consistency and integrity. Final artifacts were then transformed into early-stage entity-relationship (ER) models.

\subsection{Donor UA Case Study}

The Donor UA case study explores how \ONION can be applied to model a blood donation support system in Ukraine---a context shaped by medical stereotypes, logistical barriers, and limited digital infrastructure. Donor UA\footnote{\url{https://www.donor.ua/en}} is a platform that aims to simplify, demystify, and digitize the blood donation process by connecting donors, volunteers, and blood banks. Although a version of the platform already exists, our workshop simulated the co-design of an alternative version. Participants included experienced donors, non-donors, volunteers, and Donor UA representatives, offering perspectives from various points of interaction with the system.

This workshop was conducted online, with activities structured around the \ONION stages and supported by collaborative digital tools such as Excalidraw and Google Meet. The modeling process revealed key roles and flows---such as the ``blood journey'' and ``information journey''---which enabled participants to frame data entities and relationships as dynamic rather than static. This framing supported internal validation by ensuring that the sketched flows aligned with the semantics of the final model.

This workshop included 6 participants. To manage group size and power dynamics, we split the workshop into two teams, with 3 participants in each. Team 1 created a balanced, community-focused model through creative engagement, while Team 2 expanded the scope to include government and regulatory bodies, emphasizing system complexity and policy needs. Both teams highlighted different aspects of the same sociotechnical setting, demonstrating the adaptability of \ONION to diverse stakeholder perspectives.

\subsection{Volunteers Lab Case Study}
This on-site workshop explored a long-standing volunteer coordination system at the Ukrainian Catholic University, which lacks a digital recruitment platform. Participants included three individuals: two representatives from the Volunteers Lab and a facilitator---the first author of this paper---who had no prior expertise in the case study but was familiar with the \ONION methodology. Analog materials such as paper, markers, and stickers supported collaborative sketching in a less formal setting.

The workshop evolved from an initial focus on tracking to a broader systemic model involving volunteer lifecycle transitions and institutional support. A key moment of validation was mapping the ``student journey''---from volunteer participant to volunteer moderator. The discussion expanded into a representation of external--internal dynamics within the university ecosystem. While not all components were directly translatable into the final ER model, the process highlighted how even ``invisible'' entities shape the system and should be considered early in data model design.

\subsection{Summary of Case Insights}
Through structured facilitation, attention to group dynamics, and iterative validation, the workshops demonstrated the value of collaborative modeling in capturing both the internal logic and external influences of complex systems. These case studies showed how \ONION supports progressive abstraction from real-world narratives to technical models while maintaining participatory integrity. See supplementary materials for additional details.

\section{Discussion}
\label{sec:disc}

\begin{figure}[t!]
    \centering
\includegraphics[width=\linewidth]{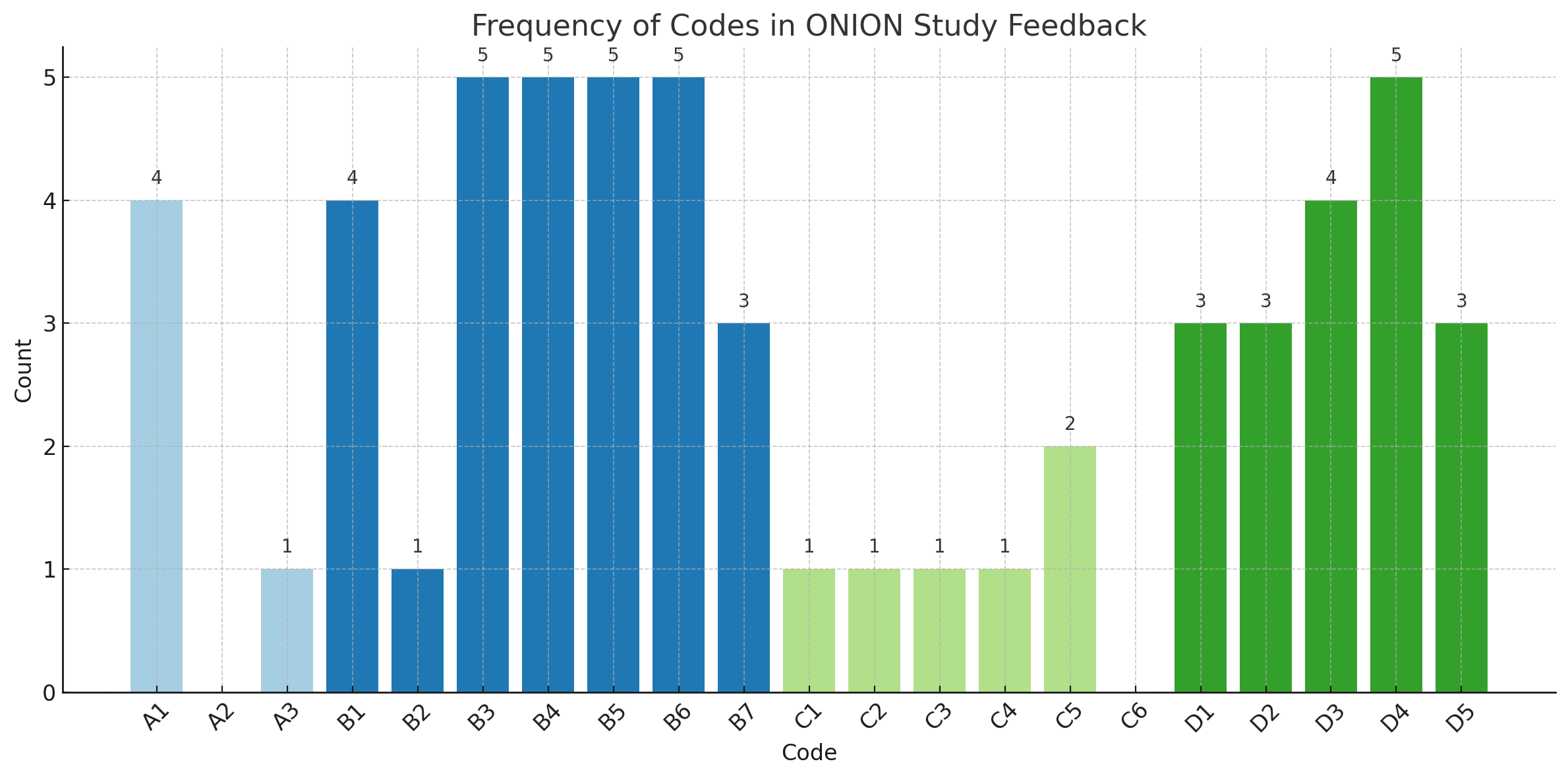}
    \caption{Frequency of codes in \ONION study feedback}
    \label{fig:codes}
\end{figure}

Our application of the \ONION framework across listed participatory workshops revealed that the methodology supports structured abstraction while maintaining stakeholder inclusion throughout the design process. We coded and analyzed participant feedback across several dimensions (see Table \ref{tab:codebook}), resulting in a set of insights that map closely to our three research questions. Figure \ref{fig:codes} shows the distribution of identified codes across all responses, helping to visualize recurring themes. More detailed description of themes and findings from participant feedbacks is presented in Table \ref{tab:rq-full-codes}.

A total of 5 participants provided feedback. Of the 6 participants in the Donor UA case study, 3 responded, while both participants in the Volunteer Lab case study gave feedback.

\begin{table}[t!] 
\caption{Summary of the codebook.}
\centering
\footnotesize

\resizebox{\linewidth}{!}{%
\begin{tabular}{|p{0.6cm}|p{5.5cm}|}
\hline
\textbf{Code} & \textbf{Explanation}\\
\hline
\textbf{A1} & No prior data modeling experience\\
\textbf{A2} & Some theoretical understanding of modeling\\
\textbf{A3} & Prior practical experience in data modeling\\
\hline
\textbf{B1} & Easy to understand and use the methodology\\
\textbf{B2} & Needed more onboarding or clearer introduction\\
\textbf{B3} & Visualization supported understanding\\
\textbf{B4} & Participants felt their input was welcomed\\
\textbf{B5} & Participants felt free to express ideas\\
\textbf{B6} & Diversity of participants improved discussion\\
\textbf{B7} & Helped participants understand ER modeling\\
\hline
\textbf{C1} & Suggestion to add more examples or templates\\
\textbf{C2} & Suggestion to clarify goals or process early on\\
\textbf{C3} & Suggests adapting to different experience levels\\
\textbf{C4} & Technical feedback on the online platform\\
\textbf{C5} & Improve onboarding/tutorial materials\\
\textbf{C6} & Time format or structure adjustment suggestions\\
\hline
\textbf{D1} & Participants experienced perspective shift\\
\textbf{D2} & Greater empathy gained for other roles\\
\textbf{D3} & Inclusion perceived as leading to fairness\\
\textbf{D4} & Final model reflects diverse input\\
\textbf{D5} & Participants realized hidden complexity\\
\hline
\end{tabular}
}

\label{tab:codebook}
\end{table}

\subsubsection*{RQ1: Methodology Effectiveness – Usability, Engagement, and Expression}

Participants generally found \ONION to be intuitive and accessible, even for those with no prior experience in data modeling. Four out of five respondents described the process as easy to understand (B1), and all mentioned that the use of live visualizations improved their comprehension (B3). While initial confusion was noted (B2), participants reported that understanding developed progressively throughout the session, confirming \ONION’s alignment with the idea of live and evolving process.

Participants noted that the environment felt safe and encouraged open sharing (B4, B5). The dialog-driven structure and supportive visualizations helped them contribute meaningfully, especially when group dynamics were balanced. However, some suggested improving onboarding--—particularly in digital settings—--by providing additional visual examples (C2, C5).

\subsubsection*{RQ2: Technical Usability and Transformation of Artifacts}

\ONION's layered structure enabled a smooth transformation of qualitative insights into structured models. Participants were able to trace how their contributions were integrated into the final ER diagrams (B7). Visualization and story-based activities played a key role in mapping early-stage insights into design artifacts. The internal validation process---particularly the use of the ``information journey'' metaphor---supported participants in evaluating whether their ideas were retained throughout the abstraction process.

Although the ER models were largely considered understandable, participants suggested improving clarity of specific elements such as attribute naming or ambiguous groupings, which were addressed during final model refinement.

\subsubsection*{RQ3: Broader Impact – Inclusion, Learning, and Bias Mitigation}

\ONION also facilitated the inclusion of diverse perspectives, which participants felt improved the final outcome (D3, D4). Many emphasized the value of engaging with individuals outside their typical roles (D2), particularly in the Donor UA case, where foundational questions were raised by non-donors. This revealed how familiar system elements may be unclear, unintuitive, or shaped by strong blood donation biases and stereotypes.

\ONION encouraged perspective shifts (D1), and several participants acknowledged gaining new insights through collaborative reflection. Even when certain ideas did not directly translate into the ER model, the recognition of hidden external influences was considered critical to ensure fairness and robustness (D5).

\begin{table}[t!]
\centering
\caption{Mapping of research questions, themes, participant feedback codes, and \ONION stages}
\small
\resizebox{\linewidth}{!}{%
\begin{tabular}{|p{0.4cm}|p{4.5cm}|p{0.8cm}|p{1.2cm}|}
\hline
\textbf{RQ} & \textbf{Theme} & \textbf{Codes} & \textbf{Stages} \\
\hline

\textbf{RQ1} & Ease of use and participant engagement. Visualization supported understanding; input was valued and free-form. & B1, B3, B4, B5, C4 & Observe, Nurture \\

\textbf{RQ1} & Initial confusion and onboarding were issues for some; simpler orientation could help. & B2, C1, C2, C5 & Observe \\

\textbf{RQ1} & Inclusion of newcomers brought diverse perspectives and boosted participant confidence. & B6, D2 & Observe, Nurture \\

\hline

\textbf{RQ2} & Participants could track how their input evolved into models; ER model mostly clear, with some unclear sections. & B7 & Integrate, Optimize \\

\textbf{RQ2} & Visual tools helped summarize stories and bridge technical concepts. & B3 & Nurture, Integrate \\

\hline

\textbf{RQ3} & Participants reported shifts in understanding and empathy through discussion with different roles. & D1, D2, D5 & Observe, Nurture \\

\textbf{RQ3} & Including diverse roles improved fairness and reduced bias in the final model.
 & C3, D3, D4 & Nurture, Integrate \\

\textbf{RQ3} & Contributions from varied backgrounds were preserved in the final model. & D4 & Integrate, Normalize \\

\hline

\textbf{BG} & Participant background: most had no prior modeling experience. & A1, A2, A3 & Observe, Normalize \\

\hline

\end{tabular}
}

\label{tab:rq-full-codes}
\end{table}

\paragraph{Practical Challenges}

\ONION demonstrates strong potential for inclusive and participatory ER modeling but also presents several practical challenges.

First, in online workshops with more than 4–5 participants, a single facilitator may struggle to manage group dynamics, ensure inclusivity, and support all participants equally. To address this, we considered splitting sessions into smaller groups.

Second, some participants produced highly complex conceptual models that extended beyond the intended ER scope. While this reflects strong engagement and systems thinking, it also underscored the need for clearer modeling boundaries or post-workshop translation to generate technically valid models. We view this as a strength, as it revealed the importance of 'invisible' entities—elements that shape the system but often go unrepresented.

Third, in digital settings, some participants were hesitant or slow to use visual tools which slowed the process and required additional facilitation to bridge verbal and visual contributions.

Finally, effective facilitation with \ONION demands skills in conflict mediation, group dynamics, and adaptive communication. Without targeted training, this requirement may limit the framework’s scalability across diverse contexts.

\section{Conclusions and Future Work}
\label{sec:conc}

The \ONION framework demonstrated its value in creating a space for mutual understanding and inclusive collaboration during early-stage data modeling. Even in small-scale settings, it helped uncover gaps in stakeholder knowledge and enabled participants to co-create meaningful models. A key contribution is \ONION’s ability to guide participants through narrowing stages while allowing creative expression.

Looking forward, the model shows promise for broader promotion and application of participatory data modeling. 
Future adaptations should also consider participants’ diverse backgrounds to ensure balanced input and inclusive design outcomes. 
Future work can also develop approaches for scaling \ONION, working more effectively with invisible entities, improved tooling and facilitator training, and porting our methodology 
to other modeling languages. 
\section*{Acknowledgments}

We gratefully acknowledge Donor UA and the Volunteers Lab at the Ukrainian Catholic University for their collaboration, participant recruitment support, and invaluable contributions to the participatory modeling workshops. We are also thankful to Anastasiia Kindzerska from Igor Sikorsky Kyiv Polytechnic Institute and Roman Petrov from  Pryazovskyi State Technical University for collaboration during early stages of research.  This work is part of the RAI for Ukraine research program, supported by the NYU Center for Responsible AI (\url{https://r-ai.co/ukraine}).

This research was conducted as part of the \href{https://r-ai.co/ukraine}{RAI for Ukraine} program, run by the Center for Responsible AI at New York University in collaboration with Ukrainian Catholic University in Lviv.   This work was supported in part by a grant from the Simons Foundation (SFARI award \#1280457, JS), and by NSF Awards No. 2326193 and 1922658.

\bibliographystyle{ACM-Reference-Format}
\bibliography{list-of-lit}


\begin{thebibliography}{26}


\ifx \showCODEN    \undefined \def \showCODEN     #1{\unskip}     \fi
\ifx \showDOI      \undefined \def \showDOI       #1{#1}\fi
\ifx \showISBNx    \undefined \def \showISBNx     #1{\unskip}     \fi
\ifx \showISBNxiii \undefined \def \showISBNxiii  #1{\unskip}     \fi
\ifx \showISSN     \undefined \def \showISSN      #1{\unskip}     \fi
\ifx \showLCCN     \undefined \def \showLCCN      #1{\unskip}     \fi
\ifx \shownote     \undefined \def \shownote      #1{#1}          \fi
\ifx \showarticletitle \undefined \def \showarticletitle #1{#1}   \fi
\ifx \showURL      \undefined \def \showURL       {\relax}        \fi
\providecommand\bibfield[2]{#2}
\providecommand\bibinfo[2]{#2}
\providecommand\natexlab[1]{#1}
\providecommand\showeprint[2][]{arXiv:#2}

\bibitem[Birhane et~al\mbox{.}(2022)]%
        {birhane2022power}
\bibfield{author}{\bibinfo{person}{Abeba Birhane}, \bibinfo{person}{William Isaac}, \bibinfo{person}{Vinodkumar Prabhakaran}, \bibinfo{person}{Mark Diaz}, \bibinfo{person}{Madeleine~Clare Elish}, \bibinfo{person}{Iason Gabriel}, {and} \bibinfo{person}{Shakir Mohamed}.} \bibinfo{year}{2022}\natexlab{}.
\newblock \showarticletitle{Power to the People? Opportunities and Challenges for Participatory {AI}}. In \bibinfo{booktitle}{\emph{EAAMO '22: Proceedings of the 2nd ACM Conference on Equity and Access in Algorithms, Mechanisms, and Optimization}}. \bibinfo{pages}{1--8}.
\newblock
\urldef\tempurl%
\url{https://doi.org/10.1145/3551624.3555290}
\showDOI{\tempurl}
\newblock
\shownote{Article No. 6}.


\bibitem[Choi et~al\mbox{.}(2020)]%
        {choi2020situated}
\bibfield{author}{\bibinfo{person}{{Jaz Hee-jeong} Choi}, \bibinfo{person}{Laura Forlano}, {and} \bibinfo{person}{Denisa {Reshef Kera}}.} \bibinfo{year}{2020}\natexlab{}.
\newblock \showarticletitle{Situated Automation: Algorithmic Creatures in Participatory Design}. In \bibinfo{booktitle}{\emph{PDC '20: Participatory Design Conference 2020 - Participation Otherwise}}.
\newblock
\urldef\tempurl%
\url{https://doi.org/10.1145/3384772.3385153}
\showDOI{\tempurl}


\bibitem[Costanza-Chock(2020)]%
        {costanza2020design}
\bibfield{author}{\bibinfo{person}{Sasha Costanza-Chock}.} \bibinfo{year}{2020}\natexlab{}.
\newblock \showarticletitle{Design Practices: “Nothing about Us without Us”}.
\newblock In \bibinfo{booktitle}{\emph{Design Justice: Community-Led Practices to Build the Worlds We Need}}. \bibinfo{publisher}{The MIT Press}, Chapter~2.
\newblock
\showISBNx{9780262356862}
\urldef\tempurl%
\url{https://doi.org/10.7551/mitpress/12255.003.0006}
\showDOI{\tempurl}


\bibitem[Dean et~al\mbox{.}(1997)]%
        {group_tooks_data_model_1997}
\bibfield{author}{\bibinfo{person}{D.L. Dean}, \bibinfo{person}{J.D. Lee}, {and} \bibinfo{person}{J.F. Nunarnaker}.} \bibinfo{year}{1997}\natexlab{}.
\newblock \showarticletitle{Group tools and methods to support data model development, standardization, and review}. In \bibinfo{booktitle}{\emph{Proceedings of the Thirtieth Hawaii International Conference on System Sciences}}, Vol.~\bibinfo{volume}{2}. \bibinfo{pages}{386--395 vol.2}.
\newblock
\urldef\tempurl%
\url{https://doi.org/10.1109/HICSS.1997.665608}
\showDOI{\tempurl}


\bibitem[Dean et~al\mbox{.}(1994)]%
        {dean1994technological}
\bibfield{author}{\bibinfo{person}{Douglas~L. Dean}, \bibinfo{person}{James~D. Lee}, \bibinfo{person}{Richard~E. Orwig}, {and} \bibinfo{person}{Douglas~R. Vogel}.} \bibinfo{year}{1994}\natexlab{}.
\newblock \showarticletitle{Technological Support for Group Process Modeling}.
\newblock \bibinfo{journal}{\emph{Journal of Management Information Systems}}  \bibinfo{volume}{11} (\bibinfo{year}{1994}), \bibinfo{pages}{43--63}.
\newblock


\bibitem[D'Ignazio and Klein(2020)]%
        {dignazio2020data}
\bibfield{author}{\bibinfo{person}{Catherine D'Ignazio} {and} \bibinfo{person}{Lauren~F Klein}.} \bibinfo{year}{2020}\natexlab{}.
\newblock \bibinfo{booktitle}{\emph{Data Feminism}}.
\newblock \bibinfo{publisher}{MIT Press}.
\newblock


\bibitem[Feinberg(2017)]%
        {Feinberg17}
\bibfield{author}{\bibinfo{person}{Melanie Feinberg}.} \bibinfo{year}{2017}\natexlab{}.
\newblock \showarticletitle{A design perspective on data}. In \bibinfo{booktitle}{\emph{{CHI}}}. \bibinfo{publisher}{{ACM}}, \bibinfo{pages}{2952--2963}.
\newblock


\bibitem[Garrett(2010)]%
        {garrett2010elements}
\bibfield{author}{\bibinfo{person}{Jesse~James Garrett}.} \bibinfo{year}{2010}\natexlab{}.
\newblock \bibinfo{booktitle}{\emph{The Elements of User Experience: User-Centered Design for the Web and Beyond}}.
\newblock \bibinfo{publisher}{New Riders}.
\newblock


\bibitem[Gero(1990)]%
        {gero1990design}
\bibfield{author}{\bibinfo{person}{John~S. Gero}.} \bibinfo{year}{1990}\natexlab{}.
\newblock \showarticletitle{Design prototypes: a knowledge representation schema for design}.
\newblock \bibinfo{journal}{\emph{AI magazine}} \bibinfo{volume}{11}, \bibinfo{number}{4} (\bibinfo{year}{1990}), \bibinfo{pages}{26--36}.
\newblock


\bibitem[Hocker et~al\mbox{.}(2020)]%
        {hocker2020participatory}
\bibfield{author}{\bibinfo{person}{Julian Hocker} {et~al\mbox{.}}} \bibinfo{year}{2020}\natexlab{}.
\newblock \showarticletitle{Participatory design for ontologies: a case study of an open science ontology for qualitative coding schemas}.
\newblock \bibinfo{journal}{\emph{Aslib Journal of Information Management}} (\bibinfo{year}{2020}).
\newblock
\newblock
\shownote{Cited by: 8}.


\bibitem[Kingsun and Hardy(2018)]%
        {kingsun2018cdom}
\bibfield{author}{\bibinfo{person}{Melinda Kingsun} {and} \bibinfo{person}{Dianna Hardy}.} \bibinfo{year}{2018}\natexlab{}.
\newblock \showarticletitle{C-DOM: a structured co-design framework methodology for ontology design and development}. In \bibinfo{booktitle}{\emph{Proceedings of the Australasian Computer Science Week Multiconference}}.
\newblock


\bibitem[Kyng(1991)]%
        {kyng1991designing}
\bibfield{author}{\bibinfo{person}{Morten Kyng}.} \bibinfo{year}{1991}\natexlab{}.
\newblock \showarticletitle{Designing for cooperation: Cooperating in design}.
\newblock \bibinfo{journal}{\emph{Commun. ACM}} \bibinfo{volume}{34}, \bibinfo{number}{12} (\bibinfo{year}{1991}), \bibinfo{pages}{65--73}.
\newblock


\bibitem[Lee et~al\mbox{.}(1997)]%
        {group_tool_data_modeling_1997}
\bibfield{author}{\bibinfo{person}{James~D. Lee}, \bibinfo{person}{Douglas~L. Dean}, {and} \bibinfo{person}{Douglas~R. Vogel}.} \bibinfo{year}{1997}\natexlab{}.
\newblock \showarticletitle{Tools and methods for group data modeling: a key enabler of enterprise modeling}.
\newblock \bibinfo{journal}{\emph{SIGGROUP Bull.}} \bibinfo{volume}{18}, \bibinfo{number}{2} (\bibinfo{date}{Aug.} \bibinfo{year}{1997}), \bibinfo{pages}{59–63}.
\newblock
\showISSN{2372-7403}
\urldef\tempurl%
\url{https://doi.org/10.1145/265665.265682}
\showDOI{\tempurl}


\bibitem[Lee et~al\mbox{.}(2000)]%
        {lee2000cold}
\bibfield{author}{\bibinfo{person}{James~D. Lee}, \bibinfo{person}{Lina Zhou}, {et~al\mbox{.}}} \bibinfo{year}{2000}\natexlab{}.
\newblock \showarticletitle{ColD SPA: a tool for collaborative process model development}. In \bibinfo{booktitle}{\emph{Proceedings of the 33rd Annual Hawaii International Conference on System Sciences}}.
\newblock


\bibitem[Norman(2013)]%
        {norman2013design}
\bibfield{author}{\bibinfo{person}{Don Norman}.} \bibinfo{year}{2013}\natexlab{}.
\newblock \bibinfo{booktitle}{\emph{The Design of Everyday Things}}.
\newblock \bibinfo{publisher}{Basic Books}.
\newblock


\bibitem[Ongenae et~al\mbox{.}(2011)]%
        {ongenae2011participatory}
\bibfield{author}{\bibinfo{person}{Femke Ongenae} {et~al\mbox{.}}} \bibinfo{year}{2011}\natexlab{}.
\newblock \showarticletitle{Participatory Design of a Continuous Care Ontology - Towards a User-driven Ontology Engineering Methodology}.
\newblock \bibinfo{journal}{\emph{International Conference on Knowledge Engineering and Ontology Development}} (\bibinfo{year}{2011}).
\newblock


\bibitem[Ongenae et~al\mbox{.}(2014)]%
        {ongenae2014ontology}
\bibfield{author}{\bibinfo{person}{Femke Ongenae} {et~al\mbox{.}}} \bibinfo{year}{2014}\natexlab{}.
\newblock \showarticletitle{An ontology co-design method for the co-creation of a continuous care ontology}.
\newblock \bibinfo{journal}{\emph{Applied Ontology}} (\bibinfo{year}{2014}).
\newblock


\bibitem[Quimby and Beresford(2022)]%
        {quimby2022participatory}
\bibfield{author}{\bibinfo{person}{Barbara Quimby} {and} \bibinfo{person}{Melissa Beresford}.} \bibinfo{year}{2022}\natexlab{}.
\newblock \showarticletitle{Participatory Modeling: A Methodology for Engaging Stakeholder Knowledge and Participation in Social Science Research}.
\newblock \bibinfo{journal}{\emph{Field Methods}} \bibinfo{volume}{35}, \bibinfo{number}{1} (\bibinfo{date}{February} \bibinfo{year}{2022}).
\newblock
\urldef\tempurl%
\url{https://doi.org/10.1177/1525822X221076986}
\showDOI{\tempurl}


\bibitem[Ram and Ramesh(1998)]%
        {ram1998collaborative}
\bibfield{author}{\bibinfo{person}{Sudha Ram} {and} \bibinfo{person}{Venkata Ramesh}.} \bibinfo{year}{1998}\natexlab{}.
\newblock \showarticletitle{Collaborative conceptual schema design: a process model and prototype system}.
\newblock \bibinfo{journal}{\emph{ACM Transactions on Information Systems}} (\bibinfo{year}{1998}).
\newblock


\bibitem[Schuler and Namioka(1993)]%
        {schuler1993participatory}
\bibfield{editor}{\bibinfo{person}{Douglas Schuler} {and} \bibinfo{person}{Aki Namioka}} (Eds.). \bibinfo{year}{1993}\natexlab{}.
\newblock \bibinfo{booktitle}{\emph{Participatory Design: Principles and Practices}}.
\newblock \bibinfo{publisher}{CRC Press}, \bibinfo{address}{Hillsdale, NJ}.
\newblock


\bibitem[Silberschatz et~al\mbox{.}(2020)]%
        {silberschatz2020}
\bibfield{author}{\bibinfo{person}{Abraham Silberschatz}, \bibinfo{person}{Henry~F. Korth}, {and} \bibinfo{person}{S. Sudarshan}.} \bibinfo{year}{2020}\natexlab{}.
\newblock \bibinfo{booktitle}{\emph{Database System Concepts} (\bibinfo{edition}{7} ed.)}.
\newblock \bibinfo{publisher}{McGraw-Hill Education}, \bibinfo{address}{New York}.
\newblock
\showISBNx{9780078022159}


\bibitem[Simsion et~al\mbox{.}(2012)]%
        {simsion}
\bibfield{author}{\bibinfo{person}{Graeme Simsion}, \bibinfo{person}{Simon~K. Milton}, {and} \bibinfo{person}{Graeme Shanks}.} \bibinfo{year}{2012}\natexlab{}.
\newblock \showarticletitle{Data modeling: Description or design?}
\newblock \bibinfo{journal}{\emph{Information \& Management}} \bibinfo{volume}{49}, \bibinfo{number}{3} (\bibinfo{year}{2012}), \bibinfo{pages}{151--163}.
\newblock


\bibitem[Simsion and Witt(2005)]%
        {simsion2005data}
\bibfield{author}{\bibinfo{person}{G. Simsion} {and} \bibinfo{person}{G. Witt}.} \bibinfo{year}{2005}\natexlab{}.
\newblock \bibinfo{booktitle}{\emph{Data Modeling Essentials}}.
\newblock \bibinfo{publisher}{Elsevier Science}.
\newblock
\urldef\tempurl%
\url{https://books.google.com.ua/books?id=-MgtlAEACAAJ}
\showURL{%
\tempurl}


\bibitem[Sloane et~al\mbox{.}(2022)]%
        {sloane2022participation}
\bibfield{author}{\bibinfo{person}{Mona Sloane}, \bibinfo{person}{Emanuel Moss}, \bibinfo{person}{Olaitan Awomolo}, {and} \bibinfo{person}{Laura Forlano}.} \bibinfo{year}{2022}\natexlab{}.
\newblock \showarticletitle{Participation Is Not a Design Fix for Machine Learning}. In \bibinfo{booktitle}{\emph{EAAMO '22: Proceedings of the 2nd ACM Conference on Equity and Access in Algorithms, Mechanisms, and Optimization}}. Article \bibinfo{articleno}{1}, \bibinfo{numpages}{6}~pages.
\newblock
\urldef\tempurl%
\url{https://doi.org/10.1145/3551624.3555285}
\showDOI{\tempurl}


\bibitem[Suresh et~al\mbox{.}(2024)]%
        {suresh2024participation}
\bibfield{author}{\bibinfo{person}{Harini Suresh}, \bibinfo{person}{Emily Tseng}, \bibinfo{person}{Meg Young}, \bibinfo{person}{Mary Gray}, \bibinfo{person}{Emma Pierson}, {and} \bibinfo{person}{Karen Levy}.} \bibinfo{year}{2024}\natexlab{}.
\newblock \showarticletitle{Participation in the Age of Foundation Models}. In \bibinfo{booktitle}{\emph{FAccT '24: Proceedings of the 2024 ACM Conference on Fairness, Accountability, and Transparency}}. \bibinfo{publisher}{ACM}, \bibinfo{address}{New York, NY, USA}, \bibinfo{pages}{1609--1621}.
\newblock
\urldef\tempurl%
\url{https://doi.org/10.1145/3630106.3658992}
\showDOI{\tempurl}


\bibitem[Vasilakis et~al\mbox{.}(2019)]%
        {vasilakis2019involving}
\bibfield{author}{\bibinfo{person}{Georgios Vasilakis}, \bibinfo{person}{Haridimos Kondylakis}, \bibinfo{person}{Martin Doerr}, {and} \bibinfo{person}{Dimitris Plexousakis}.} \bibinfo{year}{2019}\natexlab{}.
\newblock \showarticletitle{Involving domain experts in conceptual modeling: Methods, tools, and challenges}. In \bibinfo{booktitle}{\emph{Proceedings of the 38th International Conference on Conceptual Modeling (ER 2019)}}. \bibinfo{publisher}{Springer}, \bibinfo{pages}{89--97}.
\newblock


\end{thebibliography}

\newpage 
\appendix
\onecolumn
\section{Free-form system diagrams created by participants}

\begin{figure}[h!]
    \centering
    \includegraphics[width=0.7\linewidth]{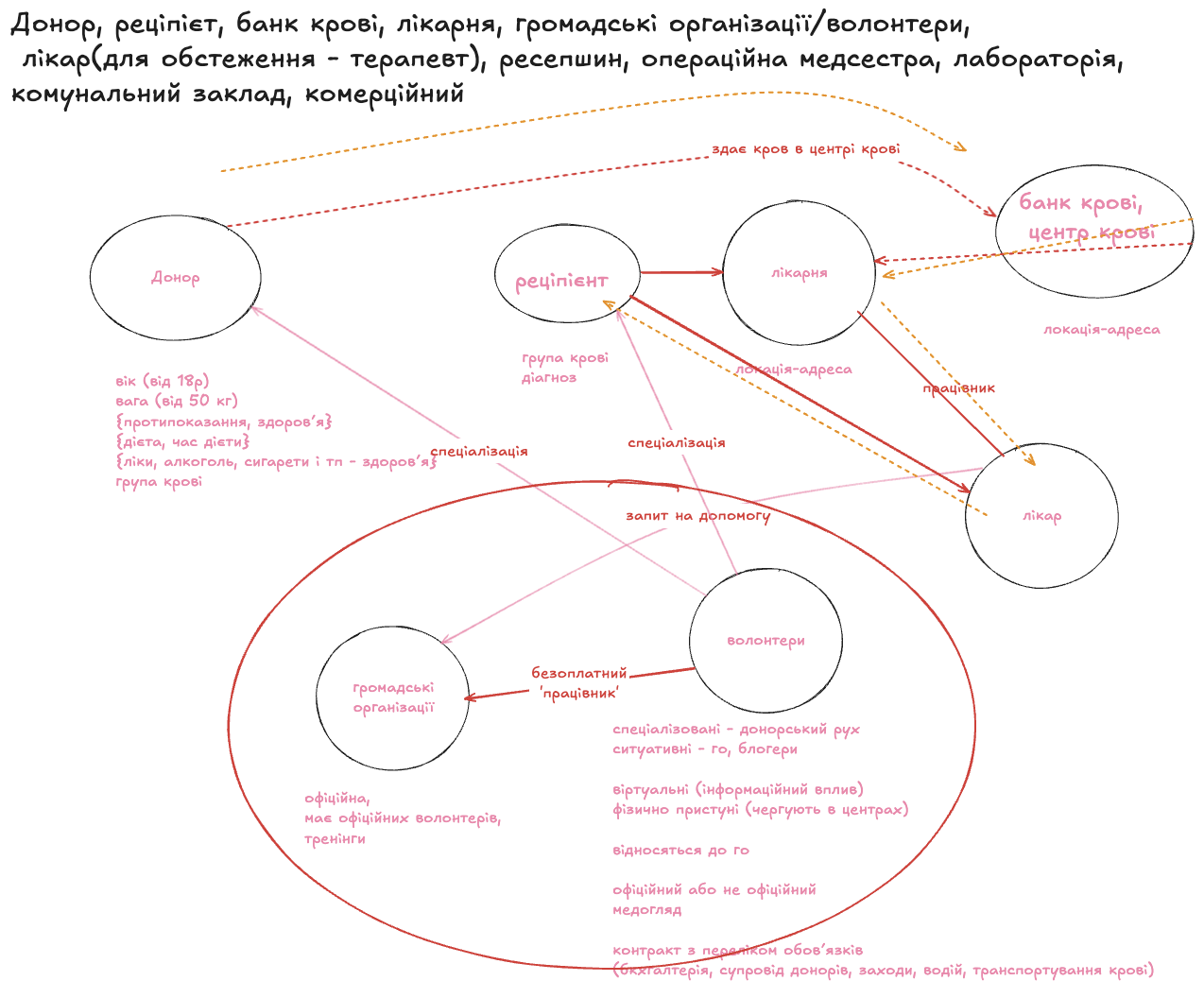}
    \caption{Free-form system diagram created by Donor UA  participants during the first workshop, with a focus on volunteers' movements.}
    \label{fig:free-form-donor-ua-1}
\end{figure}

\begin{figure}[h!]
    \centering
    \includegraphics[width=0.65\linewidth]{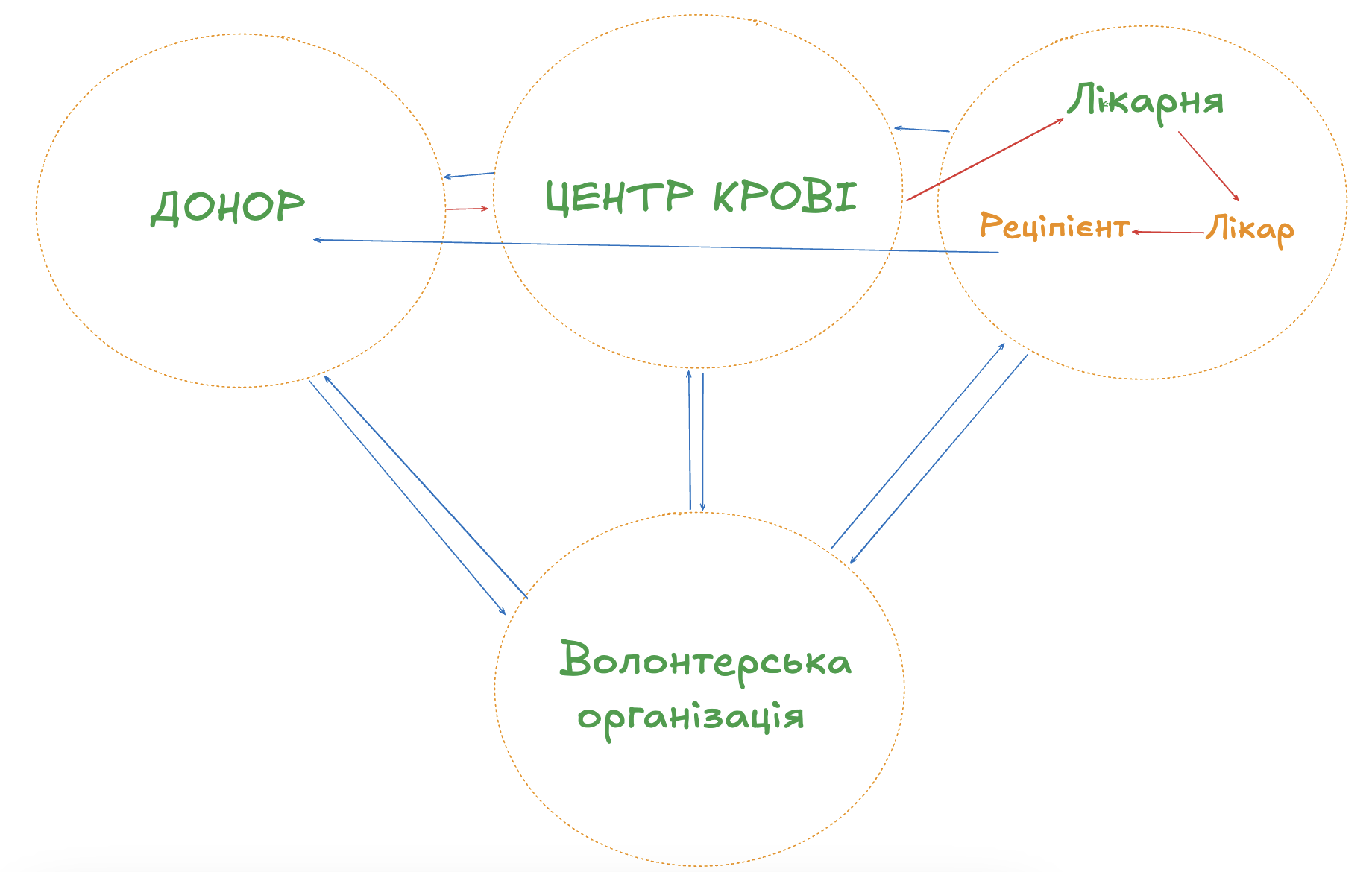}
    \caption{Free-form system diagram created by Donor UA  participants during the first workshop, highlighting the ``blood'' (red lines) and ``information'' (blue lines) paths.}
    \label{fig:free-form-donor-ua-blood-info}
\end{figure}

\begin{figure}[h!]
    \centering
    \includegraphics[width=0.65\linewidth]{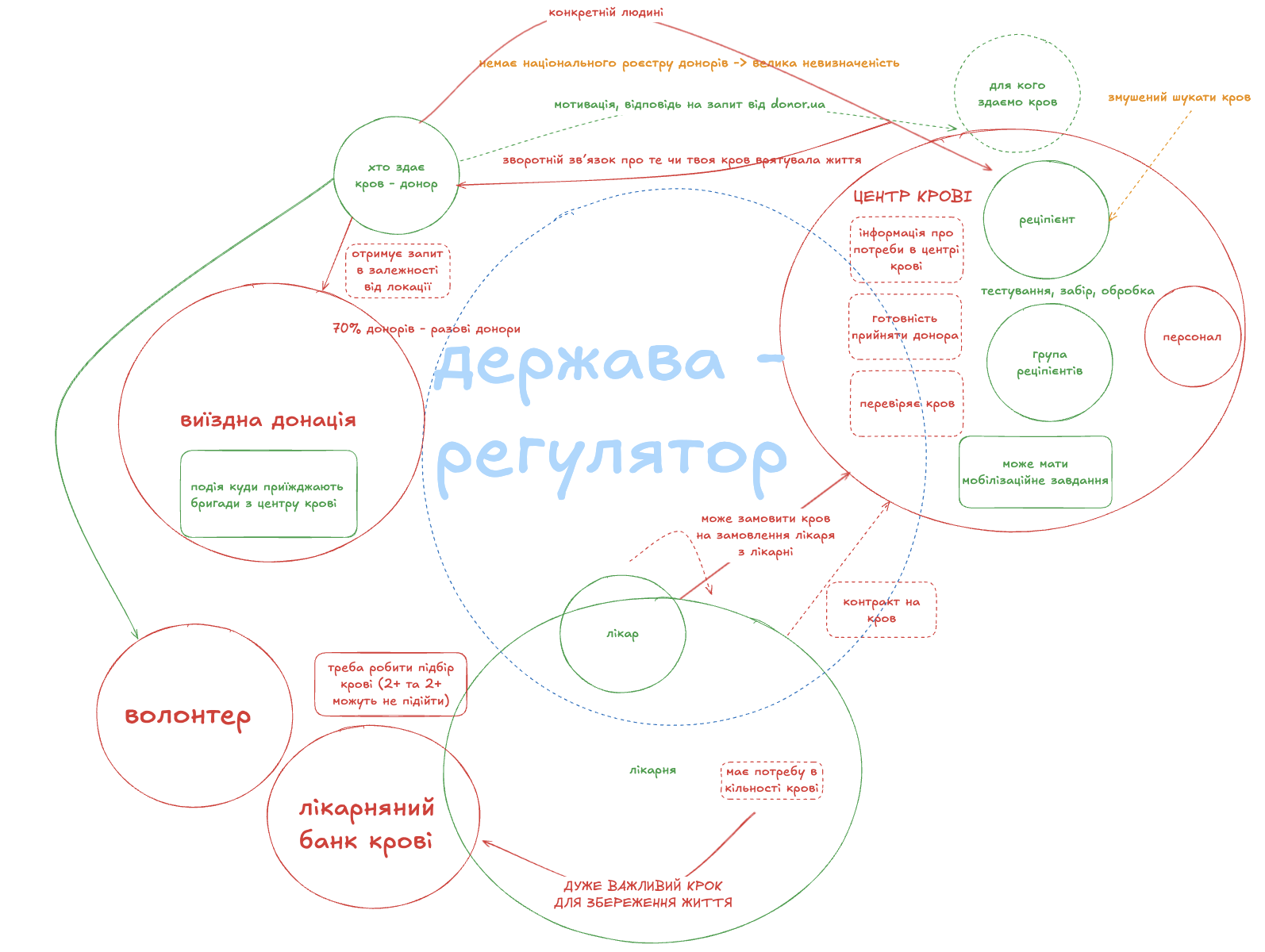}
    \caption{Free-form system diagram created by Donor UA participants during the second workshop, highlighting the role of the state and volunteers.}
    \label{fig:free-form-donor-ua-2}
\end{figure}

\begin{figure}[h!]
    \centering
    \includegraphics[width=0.68\linewidth]{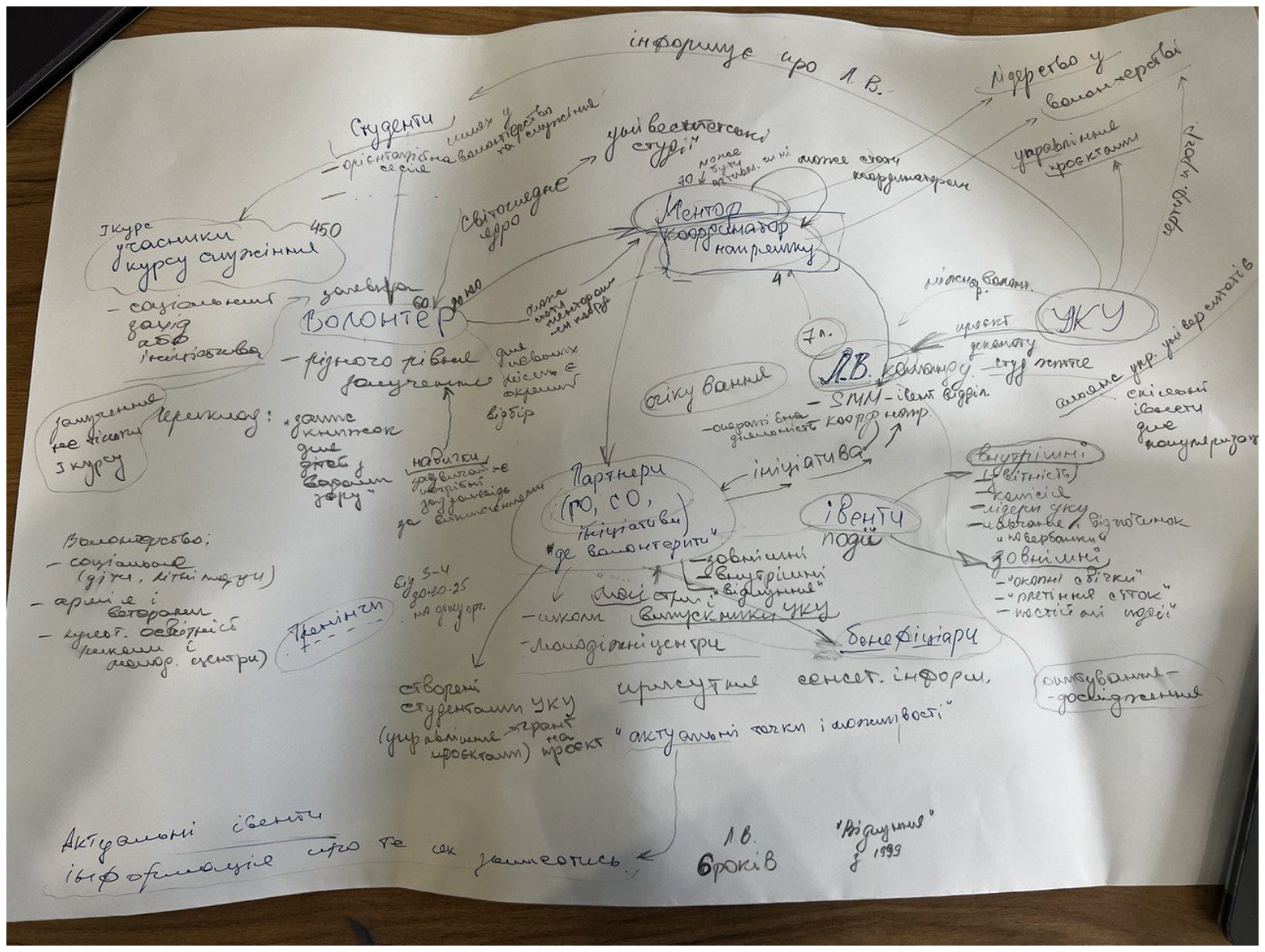}
    \caption{Free-form system diagram created by Volunteers Lab workshop participants on-site, highlighting the influence of all types of external actors in the system surrounding Volunteers Lab organization.}
    \vspace{+1cm}
    \label{fig:free-form-diagram-volunteers-lab}
\end{figure}

\newpage 

\section{ER models produced after the \ONION workshops}
\begin{figure}[h!]
    \centering
    \includegraphics[width=0.8\linewidth]{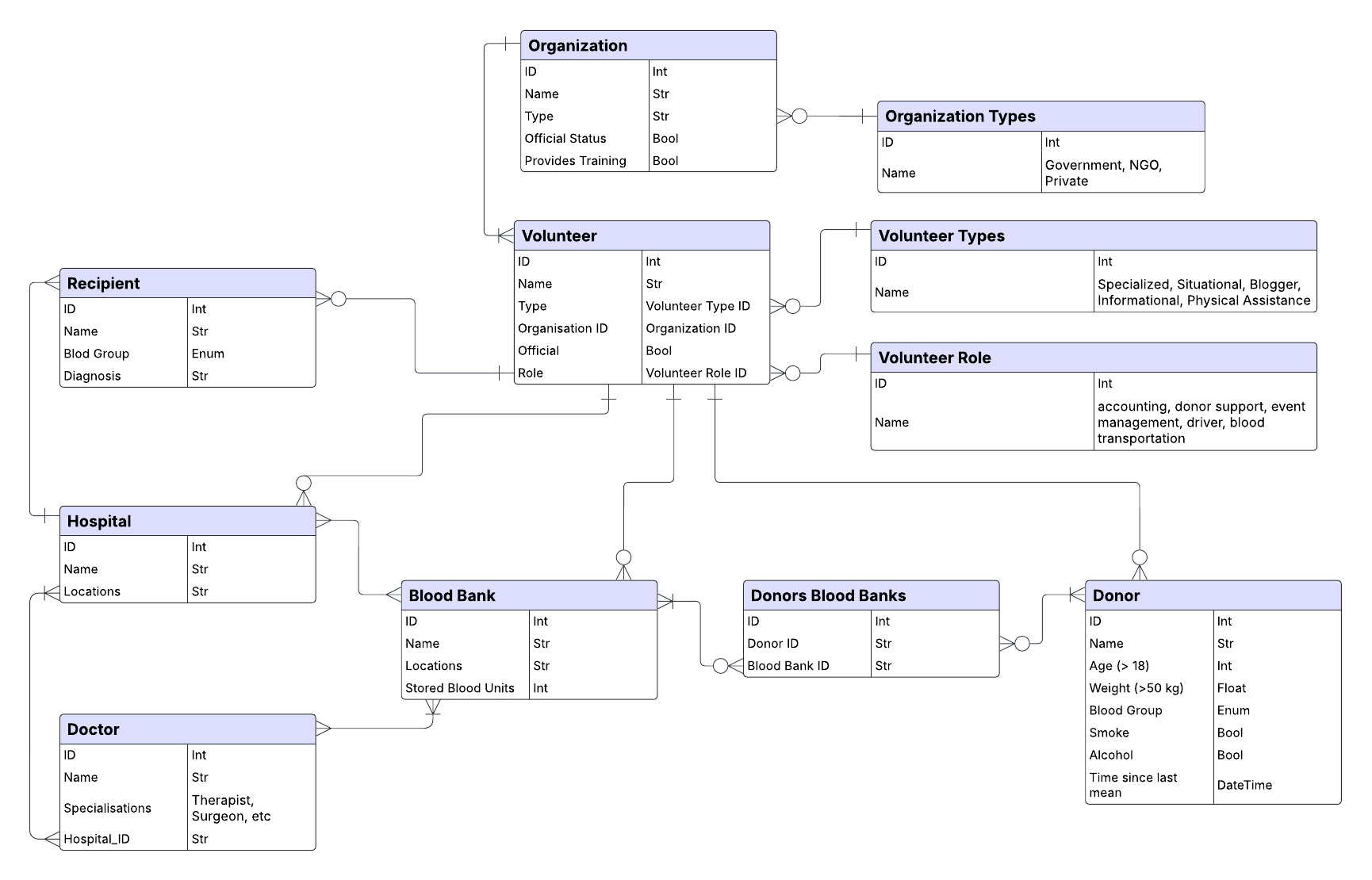}
    \caption{ER model representing the Optimize stages, created after Donor UA workshops but before the final external validation.}
    \label{fig:donor-ua-er}
\end{figure}

\begin{figure}[h!]
    \centering
    \includegraphics[width=0.8\linewidth]{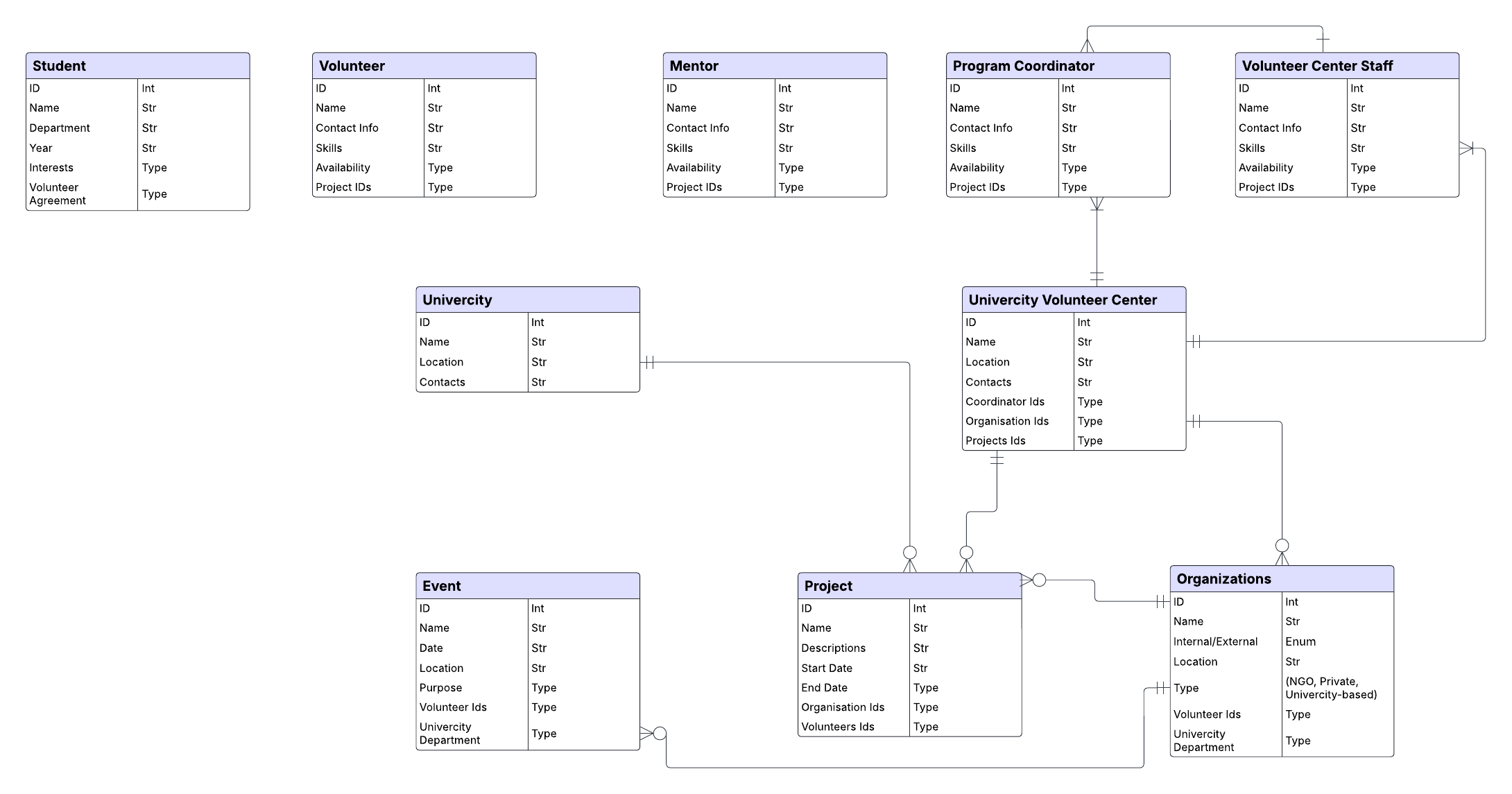}
    \caption{ER model created after Volunteers Lab workshop representing Optimize stage, which was showed to the participants to obtain their feedback before the final external validation.}
    \label{fig:volunteers-lab-er}
\end{figure} 

\end{document}